\documentclass[10pt,letterpaper]{article}
\usepackage[top=0.85in,left=2.75in,footskip=0.75in,marginparwidth=2in]{geometry}

\usepackage[utf8]{inputenc}

\usepackage{cite}

\usepackage{nameref,hyperref}

\usepackage[right]{lineno}

\usepackage{microtype}
\DisableLigatures[f]{encoding = *, family = * }

\raggedright
\usepackage{changepage}

\usepackage[aboveskip=1pt,labelfont=bf,labelsep=period,singlelinecheck=off]{caption}

\makeatletter
\renewcommand{\@biblabel}[1]{\quad#1.}
\makeatother

\usepackage{lastpage,fancyhdr,graphicx}
\usepackage{epstopdf}
\fancyheadoffset[L]{2.25in}
\fancyfootoffset[L]{2.25in}

\usepackage{color}

\definecolor{Gray}{gray}{.25}

\usepackage{graphicx}

\usepackage{sidecap}

\usepackage{wrapfig}
\usepackage[pscoord]{eso-pic}
\usepackage[fulladjust]{marginnote}
\reversemarginpar

\begin{document}
\vspace*{0.35in}

% title goes here:
\begin{flushleft}
{\Large
\textbf\newline{Meta-analysis for Discovering Which Genes\\ are Differentially Expressed in Neuroinflammation}
}
\newline
% authors go here:
\\
Berk C. Ugurdag\textsuperscript{1,*},
Serena Aktürk\textsuperscript{2},
Michelle Adams\textsuperscript{2}
\\
\bigskip
\bf{1} Robert College, Istanbul, Turkey
\\
\bf{2} Bilkent University, Ankara, Turkey
\\
\bigskip
* uguber.22@robcol.k12.tr

\end{flushleft}

\section*{Abstract}
Neuroinflammation is a significant aspect of many neurological diseases of \emph{Homo sapiens}, and the genes that are differentially expressed in this process should be well understood to gather the nature of such diseases. We have conducted a meta-analysis (based on a combined adjusted P value and logFC scheme) of 6 multi-species (\emph{Homo sapiens}, \emph{Mus musculus}) datasets (available on GEO, short for Gene Expression Omnibus) obtained through microarray technology. Our analysis shows that the genes coding pleckstrin homology domain and galectin-9 proteins take part in neuroinflammation in microglia.

% now start line numbers
% \linenumbers
% the * after section prevents numbering

\section{Introduction}
Many neurological disorders, such as Alzheimer’s and Parkinson’s diseases (PD), are found to pertain to cases of neuroinflammation, i.e., inflammation found in neurons. By disrupting the biological functions of neurons, neuroinflammation can reduce the amount of neurotransmitter secretion and activation potentials that yield neural signals, which is reflected by organismal health through behavioral and chemical changes.

Neuroinflammation involves responses of microglia, astrocytes, and oligodendroglia to certain breaches in the central nervous system. Various proteins play crucial roles in pathways that constitute neuroinflammation. Therefore, querying the effects of these proteins is of top significance in order to tackle the aforementioned exemplary diseases in future research. The shifts in the concentration of neuroinflammatory proteins require a change in protein synthesis, suggesting an increase or decrease in the transcription process, where mRNA is synthesized through cellular DNA. This difference is explained by the term differential expression.

In this study, we conducted a meta-analysis, having considered the datasets yielded by multiple scientific studies that were published in Gene Expression Omnibus (GEO) to make use of the microarray technology utilized therein, informing us about which genes have been downregulated or upregulated as a result of a series of treatments on various cells utilized in the experiments. Regarding several statistical values, the study yielded a list of significant genes. We further contextualized these genes’ biological roles through informative sources to ascertain their significance in the focused process. This further narrowed down the list. Neuroinflammation is a significant aspect of many neurological diseases of \emph{Homo sapiens}, and the genes that are differentially expressed, i.e., genes that have different impacts from different conditions in this process should be understood to gather the nature of such diseases. 

Through R programming language and GEO, we have conducted a meta-analysis (based on a combined adjusted P value and logFC scheme) of 6 datasets mostly concerning \emph{Homo sapiens} but also \emph{Mus musculus} obtained through microarray technology, which is based on the utilization of gene chips to identify the expression levels of a multitude of genes. The novel genes found in this meta-analysis will have crucial implications on which genes and proteins play a role in neuroinflammation and can be candidates as biomarkers for neuroinflammation.

\section{Datasets}
We chose datasets according to which organisms they involve and whether they focus on the inflammation of microglia, astroglia, or oligodendroglia. We specifically tried to limit the number of organisms in the study. Most datasets, therefore, involved Homo sapiens, but we also added datasets based on studies with Mus musculus. Having two mammal species as the focus of this study enabled us to better ascertain the significance and roles of the genes yielded from the upcoming statistical analysis in neuroinflammation, because in popular species used in labs like Danio rerio, most genes may not be evolutionary commonalities. 

Study A \cite{1} consequently forms a distinct autologous model of in vitro nature. Human fetal neural stem cells (NSC) were generated through adult CD34+ cells introduced to Sendai virus constructs including Oct3/4, C-MyC, Sox-2, and Klf-4. After NSC and activated autologous T cells were co-cultured or a treatment involving the recombinant of granzyme B was performed, NSC division and specialization were observed to decrease, indicating neurogenesis inhibition. 

In Study B \cite{2}, to evaluate whether antiviral genes are upregulated by the reduction of ADAR2 expression, oligodendrocytes introduced shRNA against ADAR2 or control were analyzed, using microarray technology, while the inflammatory genes and those related to immune responses were expressed more in cells that ADAR2 expression was reduced.

Study C \cite{3} includes the analysis of the transcriptomic response of human brain pericytes to 1,25-dihydroxyvitamin D3, the results of which were confirmed by RT-qPCR regarding the genes of interest. It was concluded that cultured human brain pericyte responds to 1,25-dihydroxyvitamin D3 through neuroinflammatory genes, pro-inflammatory cytokines, Interferon gamma, and TNF-alpha. Thereby, it is demonstrated that neuroinflammation is able to stimulate brain pericytes to synthesize 1,25-dihydroxyvitamin D3, while pericytes will later initiate a larger anti-inflammatory response. 

Study D \cite{4} entails the transneuronal coverage of alpha-synuclein in PD. Alpha-synuclein aggregation encouraged by the injection of alpha-synuclein preformed fibrils into mouse cerebrum was utilized. Histological analysis for alpha-synuclein inclusions, microgliosis, along with neurodegeneration in distinct regions of the brain, and a profiling of gene expression of the ventral midbrain, at two instances of time following the initiation of disease, was performed, resulting in significant degeneration in brain regions regardless of alpha-synuclein inclusions. Microgliosis independent from the portions of the brain experiencing neurodegeneration and inclusion load was noticed. Early and distinct changes were observed to be correlated with microglia-mediated inflammation before neurodegeneration in longitudinal gene expression profiling experiments, indicating active induction of neurodegeneration by microglia. Observations made show that the formation of alpha-synuclein inclusion is not the major factor during the initial stages of PD-resembling neurodegeneration. Instead, types of diffusible, oligomeric a-synuclein, which facilitate unexpectedly high levels of reaction among microglia, are considered to be the key factor in this process.

Study E \cite{5,6} weighs gene expression of human microglia and macrophages for different phases of polarization, relating these to their roles in tissue injury, defense, and reparation in the central nervous system. 

Study F \cite{7} utilizes cells originally from surgically cut tissue to show that oligodendrocytes of samples younger than the age of 5 are more resistant to in-vitro metabolic impairment in contrast to fetal O4+ progenitor cells. However, they are also found to be more vulnerable to cell demise than oligodendrocytes from adult samples. In the same study, only pediatric oligodendrocytes indicated quantifiable levels of TUNEL+ cells, which is noted to be typical to fetal cell response. The proportion of anti- to pro-apoptotic BCL-2 family genes were greater in adult compared to pediatric mature oligondendrocytes. Lysosomal gene expression turned out to be much more frequent in adult and pediatric in comparison to fetal oligodendrocyte lineage cells. Apoptosis in oligodendrocytes was also noted to be more frequent through the inhibition of pro-apoptotic BCL-2 gene, thus autophagy.

\begin{table}[!ht]
\begin{adjustwidth}{-1.5in}{0in} % comment out/remove adjustwidth environment if table fits in text column.
\centering
\caption{Dataset Information}
\begin{tabular}{|l|l|l|l|p{4cm}|}
\hline
{\bf Study} & {\bf GSE Number} & {\bf Organism} & {\bf Cell Type} & {\bf Microarray Platform} \\ \hline \hline
A & 44532 & \emph{Homo sapiens} & neural stem cells & [HuGene-2\_0-st] Affymetrix Human Gene 2.0 ST Array [transcript (gene) version] \\ \hline
B & 138927 & \emph{Homo sapiens} & oligodendroglia & [Clariom\_S\_Human] Affymetrix Clariom S Assay, Human (includes Pico Assay) \\ \hline
C & 54765 & \emph{Homo sapiens} & brain pericyte cells & [HG-U133\_Plus\_2] Affymetrix Human Genome U133 Plus 2.0 Array \\ \hline
D & 155716 & \emph{Mus Musculus} & microglia & [MoGene-2\_0-st] Affymetrix Mouse Gene 2.0 ST Array [transcript (gene) version] \\ \hline
E & 76734 & \emph{Homo sapiens} & microglia &
[HuGene-2\_0-st] Affymetrix Human Gene 2.0 ST Array [transcript (gene) version] \\ \hline
F & 160813 & \emph{Homo sapiens} & oligodendroglia & [HuGene-2\_0-st] Affymetrix Human Gene 2.0 ST Array [transcript (gene) version] \\ \hline
\end{tabular}
\label{tab1}
\end{adjustwidth}
\end{table}

\section{Methodology}
Making use of RStudio, as part of preprocessing of the datasets, we have queried GEO to download each dataset and stored its CEL type files into a variable. Normalizing the data with Robust Multi-array Average method (RMA), we have produced boxplots in order to compare differentially expressed genes in each treatment group to better ascertain whether the data was successfully normalized.

In addition, combining the GEO2R feature and our own R programs, we defined comparison groups to jot down genes that may be significant to be included in the study from the statistical values the code and GEO2R yielded: adjusted P values and fold changes. 

After picking out genes that are of interest based on significant P values (lower than 0.05) and fold changes (more than 1.5 or less than -1.5) with volcano plots,  we started converting gene IDs found in several Affymetrix versions. Our main source was database for annotation, visualization and integrated discovery (DAVID). However, this tool fell short of identifying a great number of genes yielded from the statistical analysis. Still, certain websites of probe providers assisted in the conversion process, as we were able to conserve, although small, a significant amount of unidentified genes. 

To be as accurate as possible, we noted genes as significant for further analysis based on their recurrence in a minimum of two datasets included in the study. This resulted in only 7 genes: neurobeachin-like, SNRPN, G protein, ribosomal protein, Pleckstrin homology domains, solute carrier family 22, galectin-9.

The next approach we followed was to use several free databases that provide functional information of certain genes and proteins, such as UniProt, to determine the relevance of the meager amount of remaining genes. Similar to the previous process, an overwhelming majority of genes were left out due to the fact that they were either irrelevant or too pervasive in biological functions to have a certain gravity for further inquiry.

The neurobeachin-like SNRPN, ribosomal protein, solute carrier family 22-coding genes are all too general in cellular function to be considered for neuroinflammation. However, Pleckstrin homology domains and galectin-9-coding genes have a relation with neuroinflammation, as they take part in microglial and brain activity. We suspect that G protein may be included in this list.

\begin{table}[!ht]
\begin{adjustwidth}{-1.5in}{0in} % comment out/remove adjustwidth environment if table fits in text column.
\centering
\caption{Significant Genes Recurring in Multiple Datasets}
\begin{tabular}{|l|l|l|l|l|}
\hline
{\bf Gene Name} & {\bf P Value} & {\bf LogFC} & {\bf Relevance} & {\bf Dataset} \\ \hline \hline
neurobeachin-like protein 1 (NBEAL)& 0.0009886 & 1.78 & irrelevant & A, F \\ \hline
small nuclear ribonucleoprotein polypeptide N (SNRPN) & $1\times10^-11$ & -1.70 & irrelevant & A, F \\ \hline
guanine nucleotide-binding protein (G protein) & 0.00934 & -3.34 & relevant & A, B \\ \hline
ribosomal protein (RP) & 0.0001225 & -1.92 & irrelevant & A, F \\ \hline
pleckstrin homology domains (PHIP) & 0.01092 & -4.76 & relevant &
A, B \\ \hline
solute carrier family 22 (SLC22) & 0.00934 & -3.29 & irrelevant & A, B \\ \hline
galectin-9 (LGALS9) & $4\times10^-11$ & 1.53 & relevant & A, B, E \\ \hline
\end{tabular}
\label{tab2}
\end{adjustwidth}
\end{table}

%\clearpage makes sure that all above content is printed at this point and does not invade into the upcoming content
%\clearpage

\section{Conclusion and Future Work}
We propose that the expression of pleckstrin homology domains are correlated with neuroinflammation. In fact, these proteins were previously found to play a role in immunological responses, specifically toll-like receptor-mediated immune response, against neuroinflammation, such as in PD \cite{8}.
	
Galectin-9 is a variant of $\beta$-galactoside-binding lectins, which are proteins heavily involved in neuroinflammation. Although galectin-9 has not been fully studied, there have been multiple studies that found that its expression is correlated with increased anti-inflammatory factors in microglia \cite{9,10}. We also conclude that this is the case in accordance with our analysis.
	
In addition, we believe G protein, guanine nucleotide-binding protein, may be of importance for further inquiry, since according to the tools we utilized, its role is relevant to our scope. Especially subunit gamma-2 takes part in ADORA2B-mediated anti-inflammatory cytokine production, while ADORA2B was previously found to contribute to Complete Freund’s adjuvant induced inflammatory pain \cite{11}. However, this protein is loosely relevant and is very broad and common in bodily functions, so a further study may not be fruitful at all.

Overall, we encourage the first two mentioned proteins to take part in future research, as any newly uncovered property will expose a deeper understanding in the dynamics of neuroinflammation. Therefore, these two proteins and genes coding them may be of focus for new drugs and treatments for neuroinflammation-related diseases, such as Parkinson’s disease.

%\clearpage

% \section*{Supporting Information}
% If you intend to keep supporting files separately you can do so and just provide figure captions here. Optionally make clicky links to the online file using \verb!\href{url}{description}!.

% These commands reset the figure counter and add "S" to the figure caption (e.g. "Figure S1"). This is in case you want to add actual figures and not just captions.
\setcounter{figure}{0}
\renewcommand{\thefigure}{S\arabic{figure}}

% You can use the \nameref{label} command to cite supporting items in the text.
% \subsection*{S1 Figure}
% \label{example_label}
% {\bf Caption of Figure S1.} \textbf{A}, If you want to reference supporting figures in the text, use the \verb!\nameref{}!. command. This will reference the section's heading: \nameref{example_label}.

% \subsection*{S2 Video}
% \label{example_video}
% {\bf Example Video.} Use \href{www.youtube.com}{clicky links} to the online sources of the files.

%\clearpage

% \section*{Acknowledgments}
% We thank just about everybody.

% \nolinenumbers

\section*{Acknowledgments}
This work was supported by TÜBİTAK grant 119S660.

%This is where your bibliography is generated. Make sure that your .bib file is actually called library.bib
\bibliography{library}

%This defines the bibliographies style. Search online for a list of available styles.
% \bibliographystyle{abbrv}
\bibliographystyle{ieeetr}
 
\end{document}